\documentclass[aps,prl,reprint,showpacs,preprintnumbers,superscriptaddress]{revtex4-1}
\pdfoutput=1

\usepackage{todonotes}
\usepackage{SIunits,bm,amsmath}

\renewcommand{\vec}[1]{\ensuremath{\boldsymbol{#1}}}

\begin{document}

\title{Sub-nanometer free electrons with topological charge}

\author{P. Schattschneider}
\email{schattschneider@ifp.tuwien.ac.at}
\affiliation{Institute of Solid State Physics, Vienna University of Technology, A-1040 Vienna, Austria}
\affiliation{University Service Centre for Transmission Electron Microscopy, Vienna University of Technology, A-1040 Vienna, Austria}
\author{M. St\"oger-Pollach}
\affiliation{University Service Centre for Transmission Electron Microscopy, Vienna University of Technology, A-1040 Vienna, Austria}
\author{S. L\"offler}
\affiliation{Institute for Solid State Physics, Vienna University of Technology, A-1040 Vienna, Austria}
\author{A. Steiger-Thirsfeld}
\affiliation{University Service Centre for Transmission Electron Microscopy, Vienna University of Technology, A-1040 Vienna, Austria}
\author{J. Hell}
\affiliation{Institute for Solid State Physics, Vienna University of Technology, A-1040 Vienna, Austria}
\author{J. Verbeeck}
\affiliation{EMAT, University of Antwerp, B-2020 Antwerp, Belgium}

\begin{abstract}
The holographic mask technique is used to create freely moving electrons with quantized angular momentum. With electron optical elements they can be focused to vortices with diameters below the nanometer range. The understanding of these vortex beams is important for many applications. Here we present a theory of focused free electron vortices. The agreement with experimental data is excellent. As an immediate application, fundamental experimental parameters like spherical aberration and partial coherence are determined.
\end{abstract}

\maketitle

After the observation of electrons with a vortex-like structure \cite{UchidaNature2010}, the feasibility of producing electron vortex beams with quantized angular momentum in conventional electron microscopes was demonstrated \cite{VerbeeckNature2010,McMorranScience2011} thus opening the road to applications. Vortex beams are free electrons with topological charge, characterized by a ``spiraling'' wavefront and a phase singularity at the center~\cite{PRL_v99_i19_p190404}, similar to optical vortices \cite{Bazhenov1990,Allen1992} which also carry quantized orbital angular momentum \cite{NyeBerry1974}. 
Today, there are many applications of optical vortices ranging from tweezers exerting a torque \cite{HePRL1995}, over optical micromotors~\cite{LuoAPL2000}, cooling mechanisms \cite{KuppensPRA1998}, toroidal Bose-Einstein condensates \cite{TsurumiJPSJap2001}, exoplanet detection \cite{SerabynNature2010} to quantum correlation and entanglement in many-state systems \cite{MairNature2001}.
Owing to their short wavelength, fast electrons can in principle be focused to atomic size \cite{VerbeeckScience2011}. Another aspect that makes them attractive for future applications is that---in contrast to optical vortices---they carry a magnetic moment, even for beams without spin polarization. 

The original interest in electron vortices was due to the prospective use as a filter for magnetic transitions in the ferromagnetic 3d metals \cite{VerbeeckNature2010}, thus facilitating energy loss magnetic chiral dichroism (EMCD) experiments in the electron microscope \cite{SchattNature2006}. Their actual potential is much wider, ranging from probing chiral structures to the manipulation of nanoparticles, clusters and molecules exploiting the magnetic interaction. 

The main drawback of the first investigations was twofold: i) the rather large scale of the vortex (several \micro\meter). A slightly changed scattering geometry allows the production of a series of well separated electron vortices with topological charge $|m| \ne 0$ and a beam waist of about \unit{0.2}{\nano\meter} \cite{VerbeeckScience2011}. ii) in spite of the well understood basic structure of an electron vortex, no theory of vortex propagation through a focusing optical element was available.

Here, we develop a theory that describes the characteristic features of focused electrons with topological charge. Application to sub-nanometric vortex profiles, including spherical aberration of the focusing electron optical elements and partial coherence, shows excellent agreement with experiments.

Assume a plane wave $e^{i k_z z}$ with wave number $k_z$ traveling along the optic axis of the electron microscope (in $z$ direction), which impinges onto a holographic mask with radius $R$ given by the transmission function
\begin{equation}
	T = \Pi(r/R) \cdot \frac{1}{2}(1+\cos(\vec{k}_d \cdot \vec{r}-\varphi_r))
	\label{mask}
\end{equation}
where $\varphi_r$ is the azimuthal angle and $|\vec{k}_d|=2 \pi/d$ with the lattice distance $d$ of the mask. 
The aperture is defined by the radial step function 
\[
	\Pi(r/R)=\left\{\begin{array}{ll} 1, & r\leq R \\
	0, & r>R\end{array}\right.
\]

The exit wave function is then \cite{Schattschneider2011}
\begin{multline}
	\psi = T e^{i k_z z} = 
	e^{i k_z z}\left[\frac{1}{2}\Pi(r/R) + \right.\\
	\left.e^{-i \vec{k}_d \cdot \vec{r}}\Pi(r/R)e^{\varphi_r} + e^{i \vec{k}_d \cdot \vec{r}}\Pi(r/R)e^{-\varphi_r}\vphantom{\frac{1}{2}}\right].
	\label{exit}
\end{multline}

It follows that --- apart from the first term in brackets which is the directly transmitted beam --- two of the partial waves emerging from the vortex mask are helical, of type
\[
	\psi_m(\vec{r}) \propto \Pi(r/R) e^{i m \varphi_r},
\]
where $m$ is commonly referred to as ``topological charge'' (with $m=1$ in the present example).

The phase factor $e^{\pm i \vec{k}_d \cdot \vec{r}+k_z z}$ describes the propagation of the partial waves under the angle $\pm \arctan(k_d/k_z)$ with respect to the optic axis caused by diffraction on the periodic mask \footnote{One can compensate this sideways deviation for one of the side bands by imposing a suitable initial phase $\mp k_x x$ onto the incident plane wave.}.

With a real holographic mask, it is impossible to achieve a continuous transmission functions as described in Eq.~\ref{mask}. Instead, the mask has either no transmission or complete transmission. Such a discretized mask only results in additional vortex beams with $|m| > 1$, however~\cite{VerbeeckNature2010,Schattschneider2011}.

\begin{figure}
	\centering
		\includegraphics[height=5cm]{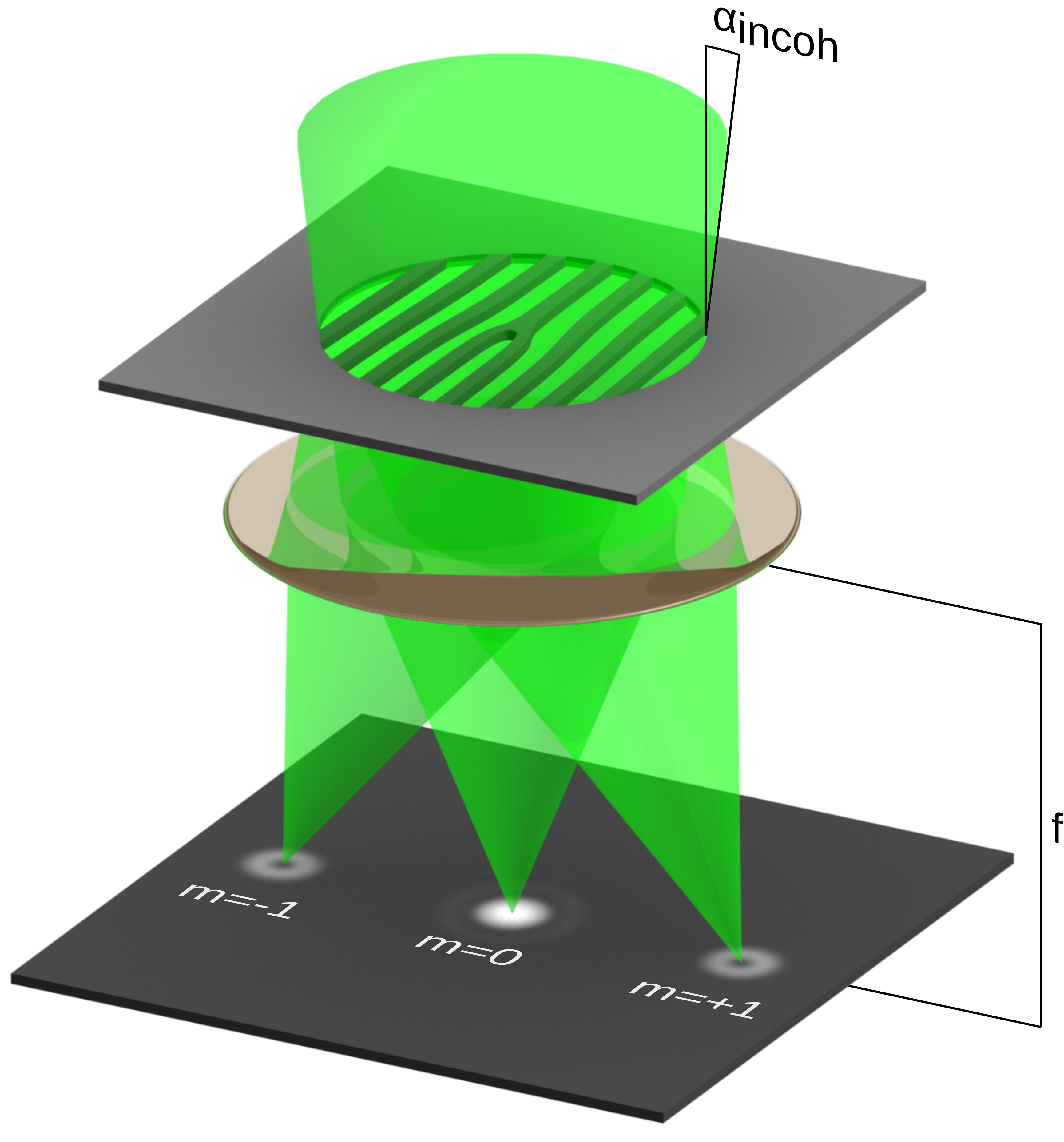}
	\caption{Sketch of the experimental setup (not to scale). For simplicity, only the essential components are drawn. From the top, the electron beam hits the aperture with the screw dislocation (gray). There, it is diffracted into the central beam as well as side bands. These are focused by a lens (brown) on the object plane (black), where the typical donut shapes ($|m| \ge  1$) and Airy disk ($m=0$) are well separated.}
	\label{fig:setup}
\end{figure}

The electron optical illumination system focuses the exit wave of the mask onto the object plane. There, we find a triple set of beam waists, well separated when $d \ll R$.
The beams focused on the object plane are given by the Fourier transform of the exit wave function Eq.~\ref{exit}. Fig.\ref{fig:setup} shows the setup with the mask and the recorded arrangement of electron vortices; the central minima for $|m|\neq 0$ are clearly visible.

According to a theorem for the Fourier-Bessel transform of a function $f(r)e^{im \varphi_r}$,
the Fourier transform of $\psi_m$ is
\begin{equation}
\tilde \psi_m(\vec{x}) = \frac{i^{m}}{2 \pi} e^{im\varphi_x}\int_0^{R} J_{m}(x r) r dr,
\label{FT2}
\end{equation}
where $\varphi_x$ is now the azimuth angle in the object plane. Via the shift theorem one finds a superposition of wave functions
\begin{equation}
\tilde \psi(\vec{x}) = e^{i k_z z_0} \sum_{m=-1}^1\tilde \psi_m(\vec{x}-m \vec{k}_d)
\label{FT3}
\end{equation}
in the object plane (arbitrarily set at a $z=z_0$). When the vortices do not overlap, the intensity $I_m(q) = |\tilde\psi_m|^2$ of each vortex is isotropic. Only the $m=0$ beam has a maximum at the center; it is the well known Airy disk. The helical beams are characterized by a ringlike structure with a radius depending on the helicity $m$.

\begin{figure}
	\centering
	\includegraphics{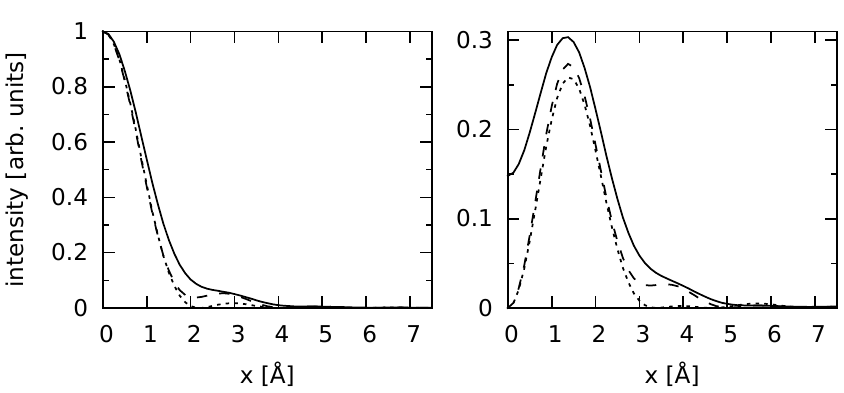}
	\caption{Profiles without spherical abberations and partial incoherence effects (dotted), with $C_s = \unit{1.2}{\milli\meter}$ (dashed), and with $C_s = \unit{1.2}{\milli\meter}$ and $\alpha_\text{incoh} = \unit{13}{\nano\rad}$ (full lines). $m=0$ on the left, $m=1$ on the right. A semi-convergence angle of \unit{6.9}{\milli\rad} was used. All curves are normalized to the maximum of the Airy disks.}
	\label{fig:Aberr}
\end{figure}

The spherical aberration and defocus of the lens must be taken into account for the analysis of the experimental results. The lens imposes an additional phase of \cite{Reimer1984}
\begin{equation}
	\chi = \frac{df q^2}{2k_z}+\frac{C_s q^4}{4 k_z^3}
	\label{eq:chi}
\end{equation}
onto the exit wave of the mask ($df$ is the defocus value, and $q=k_z r/f$ with the focal length $f$ of the lens).
Instead of Eq.~\ref{FT2}, we therefore have
\begin{equation}
\tilde \psi_m(\vec{x})=
\frac{i^{m}}{2 \pi} e^{im\varphi_x}\int_0^{R} e^{i \chi} J_{m}(x r) r dr,
\label{FT4}
\end{equation}
which we evaluate at the beam waist \footnote{This is at the Scherzer defocus $df=-\sqrt{C_s \lambda}$.}.
Results are shown in Fig.\ref{fig:Aberr}.
Note the side maxima in the profiles with aberration, which are strong for $m>0$. Importantly, the maxima are not altered by aberration. Also the central zero for the vortices with $|m|>0$ remains because $J_m(0) = 0$ for $m \ne 0$.

In reality, a perfectly coherent probe does not exist. Instead, we have illumination from extended sources which reduce the spatial (lateral) coherence of the beam. The degree of coherence is given by the angle $\alpha_\text{incoh}$ subtended by the source on the aperture. The incoherent superposition of waves coming from different points in the source reduces the size of the coherence patch on the aperture.
The resulting vortex profile is calculated from a convolution of a coherent vortex,
\[
	\rho_m=|\tilde \psi_m(\vec{x})|^2,
\]
obtained from Eq.~\ref{FT4} with the incoherent intensity distribution of the source (as usual assumed as Gaussian), again projected onto the object plane
\[
	\rho_s=e^{-\frac{1}{2}\vec{x}^2/(\alpha_\text{incoh} f)^2},
\]
where $\alpha_\text{incoh} f$ corresponds to the projected source size.

After some algebra, the profile
\begin{equation}
I_m=\rho_s \otimes \rho_m
\label{Ix}
\end{equation}
can be written as
\begin{multline}
	I_m(x)= e^{-\frac{1}{2}x^2/(\alpha_\text{incoh} f)^2} \\
	\int_0^\infty \rho_m(r) e^{-\frac{1}{2}r^2/(\alpha_\text{incoh} f)^2} I_0( x r/(\alpha_\text{incoh} f)^2) r dr,
\end{multline}
with the modified Bessel function of first kind and order zero, $I_0$.

Fig.~\ref{fig:Aberr} shows the effect of partial coherence for a typical choice of $\alpha_\text{incoh}$. The central dip is increased from zero to half the maximum at $\alpha_\text{incoh}=\unit{13}{\nano\rad}$. This constitutes an extremely sensitive monitor of beam coherence and provides detailed information about the characteristics of the incident beam down to the picometer range (in terms of the corresponding projected source size).

\begin{figure}[htbp]
	\centering
	\includegraphics{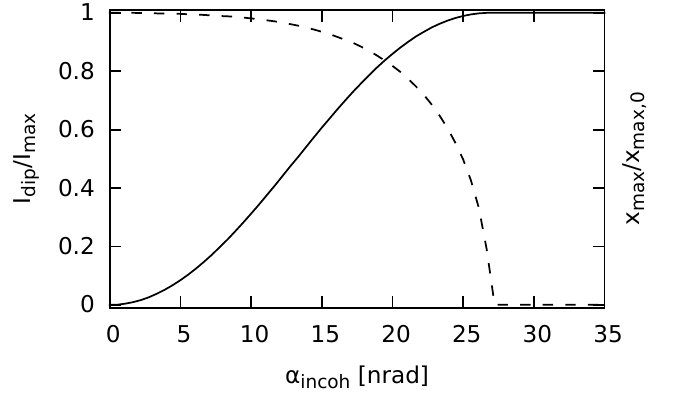}
	\caption{Central dip (full line) of the $m=1$ vortex as function of the incoherence angle $\alpha_\text{incoh}$ in the object plane. Position of the donut maximum $x_\text{max}$ relative to the position of the maximum in the perfectly coherent case $x_\text{max,0}$ (dashed line). Both curves were calculated for $C_s=\unit{1.2}{\milli\meter}$ and a semi-convergence angle of \unit{6.9}{\milli\rad}.}
	\label{fig:Dip}
\end{figure}
The central dip value (relative to the maximum) as a function of $\alpha_\text{incoh}$ is shown in Fig.~\ref{fig:Dip} for the actual $C_s$ and convergence angle.

For the experimental verification of these calculations, a TECNAI F20 microscope operated at \unit{200}{\kilo\volt} was used. At this setting, the spherical aberration coefficient is $C_s=\unit{1.2}{\milli\meter}$. The C$_2$ holographic mask was produced by depositing Cr onto a conventional aperture in a sputter plant of special geometry \cite{Hell2009453}. Into this film the characteristic fork-dislocation pattern (see Fig.~\ref{fig:setup} and Ref.~\cite{VerbeeckNature2010}) was cut by a focused ion beam (FIB). The diameter of this mask aperture was \unit{50}{\micro\meter}. 
The experimental parameters correspond to a focal length of $f=\unit{3.6}{\milli\meter}$, which is equivalent to a convergence semi-angle of $\unit{6.9}{\milli\rad}$.

To obtain radial profiles and increase the signal to noise ratio, each vortex was averaged over the polar angle $\varphi_x$~\footnote{The distance between adjacent spots was more than 25 times larger than the half width at half maximum of the $m=0$ spot to ensure negligible contributions from other spots in the averaging procedure.}.

\begin{figure}[htbp]
	\centering
	\includegraphics{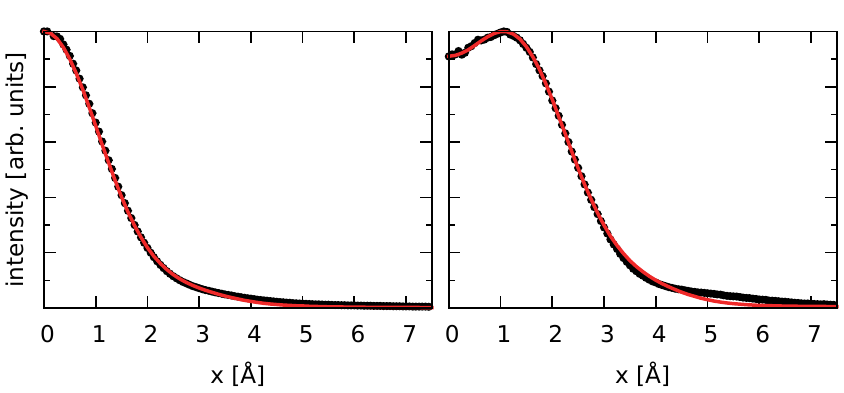}
	\caption{Experimental nanoprobe radial profiles (symbols) of vortices with $m=0$ (left) and $m=1$ (right)  compared to simulations assuming $\alpha_\text{incoh}=\unit{21.4}{\nano\rad}$, $C_s = \unit{1.2}{\milli\meter}$, and $\alpha = \unit{6.9}{\milli\rad}$.}
	\label{fig:nPsim}
\end{figure}
We measure a central dip of 0.91 for the $m=1$ vortex. From the relationship central dip--source width (Fig.~\ref{fig:Dip}) we infer $\alpha_\text{incoh}=\unit{21.4}{\nano\rad}$. Fig.~\ref{fig:nPsim} shows the radial profiles compared to the simulations with this parameter. The agreement is excellent. The minor discrepancies for $x > \unit{4}{\angstrom}$ are attributed to the fact that the electron wave incident on the mask is not an ideal plane wave due to irregularities in the gun and the (uncorrected) C$_1$ lens system.

In practice, the defocus $df$ of the condenser system is essential for data interpretation.
According to Eq.~\ref{eq:chi}, such a defocus changes the phase transfer behavior of the lens and therefore the measured intensity.

The same holographic aperture and microscope were used for the defocus series, but the latter was operated at 
an equivalent focal length of \unit{10.3}{\milli\meter} corresponding to a convergence semi-angle of \unit{2.43}{\milli\rad}. This way, the vortex radius becomes larger than a few angstroms, but both intensity and the resolution with which the vortex is imaged are improved significantly.

\begin{figure}[htbp]
	\centering
	\includegraphics{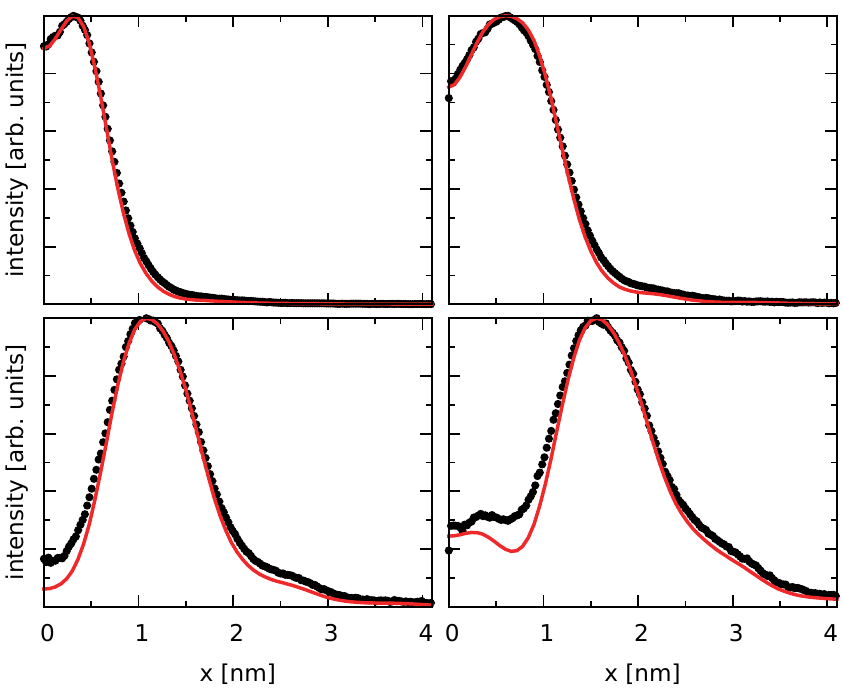}
	\caption{Experimental microprobe radial profiles (symbols) of vortices with $m=1$ compared to simulations for defocus values of \unit{200}{\nano\meter} (top left), \unit{545}{\nano\meter} (top right), \unit{890}{\nano\meter} (bottom left), and \unit{1250}{\nano\meter} (bottom right). For the calculations, the values $\alpha_\text{incoh}=\unit{21.4}{\nano\rad}$, $C_s = \unit{1.2}{\milli\meter}$, and $\alpha = \unit{2.43}{\milli\rad}$ were used.}
	\label{fig:defocus}
\end{figure}

In Fig.~\ref{fig:defocus}, experimental defocused vortices are compared to simulations \footnote{It should be noted that due to the unique properties of the condenser--objective lens system, the real defocus $df$ given here is not identical to the value displayed on the user interface. This is due to the fact that the latter is only valid for (approximately) parallel illumination \cite{Glaser1952}.}. The agreement is remarkable, considering the very large range of defoci used (up to \unit{1250}{\nano\meter}). It is clearly visible that with increasing defocus, the diameter of the vortex beams increases and the central dip is lowered. This seems to be the exact opposite to the effect of $\alpha_\text{incoh}$, but the underlying physics is very different. It must be emphasized that this lowering of the central dip does not correspond to an improvement of vortex quality or coherence but is simply a result of the Fresnel propagator that transforms the fully focused beam into the defocused one. Experimenters must be aware of this, as otherwise results would be severely misinterpreted.

The manipulation of free electrons with topological charge and a beam waist below the {\nano\meter} range 
in an electron optical system is well described within the theoretical framework developed here. The simulation of vortex shapes is surprisingly accurate. As a first application,
the possibility to derive pertinent experimental parameters such as the spherical aberration coefficient $C_s$ or the incoherence angle $a_\text{incoh}$ from the measured data was demonstrated.
The interpretation of vortex-related experiments is feasible on this basis.

\begin{acknowledgments}
PS and SL acknowledge the support of the Austrian Science Fund (FWF) under grant number I543-N20.
\end{acknowledgments}

\bibliography{NewRefs,quickbib1,schatbib}

\end{document}